# Convivial Solipsism as a maximally perspectival interpretation

**Hervé Zwirn**

Centre Borelli (ENS Paris – Saclay, 4, avenue des Sciences, 91190 Gif-sur-Yvette)

&

IHPST (CNRS, ENS Ulm, University Paris 1, 13 rue du Four, 75006 Paris, France)

herve.zwirn@gmail.com

**Abstract:** A classification of different interpretations of the quantum formalism is examined and the concept of perspectival interpretation is presented. A perspectival interpretation implies that the truth is relative to the observer. The degree to which QBism and Convivial Solipsism are perspectival is examined and Convivial Solipsim is shown to be perspectival at a higher degree than QBism or at least than the QBism founders' version.

1. Introduction

Quantum mechanics was created a century ago, and its predictions have never been disproved by experiment. Despite this, no consensus has been reached as to how quantum formalism should be interpreted. There are many competing interpretations. Establishing a taxonomy of these interpretations is tricky, as many of them have points in common and differ only in subtle details.

A first criterion to consider is whether the state vector $\Psi$ is supposed to represent the real state of the system or simply the knowledge we have of it. This is the now classical distinction between the ontic and the epistemic interpretations of the wave function [1].

The ontic view assumes that state vectors directly represent reality. The epistemic view on the contrary posits that they refer only to our knowledge of part of reality. Moreover, the quantum state can be considered as a complete description of reality - which is called the $\psi$-complete view - or may require to be supplemented with additional variables –this is the $\psi$-incomplete view [2]. The ontic $\psi$-complete view is generally regarded as the orthodox interpretation of quantum mechanics. In the $\psi$-incomplete view, it is assumed that the real state of the system, usually denoted by $\lambda$, is not totally specified by the quantum state which is compatible with several $\lambda$. So there is a probability distribution of $\lambda$ over $\psi$, $p(\lambda|\psi)$. Harrigan and Spekkens [2] noticed that if the probability distributions associated with two different quantum states $\psi$ and $\psi'$ are never overlapping, then each ontic state $\lambda$ encodes only one quantum state because one $\lambda$ is consistent with only one $\psi$. In this sense the quantum state can be considered as ontic because a variation of $\psi$ implies a variation of reality and so, even incomplete, $\psi$ captures a true aspect of reality. On the contrary, if the probability distributions associated with two different quantum states $\psi$ and $\psi'$ are overlapping, then $\psi$ and $\psi'$ are compatible at least with one ontic state. Then they can be considered as referring simply to our knowledge much in the same way as probability distributions in statistical mechanics are. This is the ψ-epistemic view. An important thing to notice is that in the ontic view, the collapse of the wave function following a measurement is a real physical change of the state of the system while in the epistemic view, it is only an update of the knowledge of the observer. Then in the latter case, the collapse is not a physical action on the system, and the proponents of this position contend that this solves the measurement problem. Nevertheless, this needs a deeper examination.

Many no-go theorems have been proved for models fitting this taxonomy. Under the assumption that systems that are prepared independently have independent physical states, Pusey et al. [3] show that if the quantum state can be specified more precisely than the ψ description with ontic states λ, then one λ cannot belong to several ψ. Then only the ψ-ontic view is correct. A stronger claim is made by Colbeck and Renner [4] who prove that under the assumptions that measurement settings can be chosen freely and that there is no superluminal signalling of the choice at the ontic level, any model must be ψ-complete. The ψ-epistemic view then seems difficult to support. Nevertheless, Lewis et al. [5] show that if the preparation independence assumption is abandoned and if superluminal influences of measurement choices on ontic variables are allowed then it is possible to build a ψ-epistemic model.

Are these results really devastating for non-ontic interpretations? No because the Harrigan and Spekkens' taxonomy is relevant mainly in a realist context such that there is "really" a physical system independent of the presence or of the knowledge of any observer and that to this system corresponds an ontic state λ, possibly identical to ψ in the complete case. This is what I call the metaphor of the theatre [6]. In this conception, the world is a kind of theatre inside which there are systems and where all the events take place. An event is a change in some property (position, momentum, energy, type of particle…) belonging to a system and taking place at a certain time and a certain location inside a given reference frame. We can witness such an event or not. This makes no difference. The events are objective facts. This picture is compatible with a fully realist position: the universe exists independently of any observer and everything that takes place inside it happens really as it is described by the theory. Observers play a passive role limited to witnessing what happens independently of them. Facts happen and are facts for the universe; they are absolute facts for all observers. So the observers are facing a scene where all the events take place. This is typically a description which fits with classical physics.

Now many instrumentalist, pragmatist or anti realist positions does not fit well in this description and are not correctly taken into account by the above classification. The no-go theorems do not apply to them and they can be epistemic in some way.

This is the case of QBism [7, 8, 9, 10] which is an epistemic interpretation (QBists prefer to say a doxastic interpretation) that does not say that the state vector is about the real world but instead is just a tool allowing any agent to compute her (subjective) probabilistic expectations for her future experience from the knowledge of the results of her past experience.

Everett's interpretation [11, 12, 13] whose initial motivation is to get rid of the reduction postulate, though realist in some way does not fit well neither in this classification. That's also the case of the Relational interpretation [14].

For this kind of epistemic interpretations that do not fit well in the Harrigan and Spekkens taxonomy, it's interesting considering what Matt Leifer [15] proposes to call "Copenhagenish" interpretations and that are closer in spirit to the ideas of Bohr, Heisenberg, Pauli et. al. than they are to the strong realism favoured by the likes of John Bell. They satisfy the four following principles:

(1) The quantum state is epistemic (information, knowledge, beliefs). It is not a direct representation of reality.
(2) Outcomes are unique for a given observer.
(3) Quantum theory is universal. It applies to every system whatever its size. In particular, observers can be described by quantum theory.
(4) Quantum theory is complete. It does not need to be supplemented by hidden variables. There is no more fundamental ontic states assigned to systems.

In this category he puts QBism [7, 8, 9, 10], Healey's pragmatism [16, 17, 18], Bruckner's position [19], Bub's and Pitowski's information-theoretic interpretation [20] and Relational Quantum Mechanics [14]. Deciding if Everett's interpretation is Copenhagenish depends on the way you interpret Everett's interpretation (there are many ways to understand it) and on what you call an observer. During a measurement the observer splits in as many copies of herself as there are possible results. If you call observer the person who exists in each world after the split then each observer sees a unique outcome and it is Copenhagenish. If on the contrary you consider that the observer is the person who makes the measurement before the split then it is not. This can be open to debates. Nonetheless there is an important difference: Everett takes the wave function as a description of the real world, in contradiction with principle 1.

These interpretations fall into two categories: objective ones and perspectival ones. The objective ones consider that the results obtained after a measurement, what observers observe, are facts about the universe. This is what I described above using the picture of the theatre. The perspectival ones consider that what is true depends on the observer. So, the result an observer gets is a result for her but not necessarily for another observer. Different people can have different truths. As Leifer says: "what is true depends on where you are sitting". Healey's interpretation and Bub's and Pitowski's interpretation are objective[1] while QBism and Relational Quantum Mechanics are perspectival[2].

The interesting point is that Leifer shows that due to what he calls the Bell/Wigner mashup no-go theorem, Copenhagenish interpretations should be perspectival. Convivial Solipsism belongs clearly to this family and is, as we will see, probably the most perspectival of all these interpretations.

---

[1] It seems nevertheless that Healey recently changed his mind [private communication].
2 See the comparison between QBism and Relational Quantum Mechanics made by Pienaar [21, 22]).

## 2. Is it enough to say that the state vector does not represent a physical state?

It is often said that quantum paradoxes are not merely solved but are dissolved if one adopts the view that the state vector is not representing the physical state of the system (ontic point of view) but the agent's knowledge or the agent's belief. In one word, that the state vector is nothing but a tool for helping to compute the probability to get such or such result. That is partially true but demands a more detailed analysis.

The point I want to insist on is that whatever the status we give to the state vector (whether it is directly representing the real physical state of the system or it represents the knowledge that the observer has of the system or the belief that the agent has in order to make future predictions or anything else) we have to take into account that it is mandatory to notice that the result that the observer gets from the act of measurement is not something pre-existing but is created during the measurement. It is obvious that certain status given to the state vector in the list above are immediately problematic. The status of directly representing the physical reality of the system leads directly to the many well-known paradoxes of the quantum formalism. The epistemic (in the broad sense) status does not face the same problems. Nevertheless we cannot be satisfied with this sole fact. To claim to have completely solved the problems associated with the formalism it remains necessary to analyse precisely what a measurement is and how it can give rise to a result.

Let's be more precise about that. In the classical realist representation that we all have of our everyday world, events occur spontaneously (it's raining, a shutter slams, a solar eclipse occurs) or are produced by an action on our part (I roll a die) but we passively observe them once they have occurred. What is more, these events are objective in the sense that they are true for any other person, they are true for the universe. Once this is established, we can have tools that allow us to predict these events either with certainty or probabilistically. Celestial mechanics allows us to predict with great precision when the next solar eclipse will occur. We can calculate the probability that a thrown die will land on 6. We can use degrees of subjective belief to predict who will win the final of a tennis tournament. But the tools we use tell us nothing about the production of events. We must clearly dissociate the event and its mode of production from the analysis of the tools we can use to predict its occurrence or probability. The observation of the event can be considered to be independent of the occurrence of the event itself. Once the event has occurred, it may (or may not) be observed. This will have no effect on the event itself. This is not the case in quantum mechanics. We know from the Kochen-Specker theorem [23] that quantum physics is contextual, which means that it is impossible to assign simultaneously defined values to all the properties of interest for a system. As a result, the usual realist view (described above), in which an observer can simply passively observe these properties (hence can use a classical probability calculation that merely reflects its ignorance), must be abandoned in favour of a view in which the system reveals its properties only when experiments are directed towards some of them to the exclusion of others, and reveals them in a non-deterministic way. An observer who makes a measurement of a property therefore performs an action that has the effect of creating the value obtained, and does so in a non-deterministic way following a contextual probability calculation. An observation or a measurement is not a passive action to witness something that is a pre-existing state of affairs but on the contrary it is an action that contributes to create the result that is observed.

Now, let's examine for example how Bitbol [24] *dissolves* the Schrödinger's cat paradox in a way that he applies to QBism. Bitbol emphasises that the state vector does not describe the state of the system but is merely a means of evaluating the probabilities of experimental results. In the Schrödinger's cat experiment, the superposition of the state vector does not therefore refer to the state of the cat but to a certain contextual modelling of the observer's uncertainty in a calculus of probabilities that uses probability amplitudes. The superposed state vector [cat dead + cat alive] does not therefore refer to a state in which the cat is both dead and alive. It is nothing more than a calculation tool from this contextual probability calculation, that must not be interpreted as such, which allows us to calculate the probability that the cat be dead or alive if we observe it. Bitbol rightly points out that the paradox raised by a cat that is both dead and alive in an interpretation where the state vector represents the real state of the system is dissolved.

But we cannot be satisfied with this sole explanation. In this discussion we need to explain both the status we attribute to the state vector and to describe what an observation is. Saying that the state vector does not directly describe the physical state of the system but is a subjective belief that only serves to calculate the probability that we attribute to what we can observe if we make a measurement is a possible solution to the apparent contradiction of describing a cat that is both alive and dead (in the cat paradox) or to that of noting that Wigner and his friend do not have the same state vector in the problem of Wigner's friend. The paradox associated with a direct realist interpretation of the state vector no longer exists, but the question of observation remains. Let's now clearly separate the "collapse of the wave-function" which is the operation of updating the tool that the state vector is in order to acknowledge the fact that a definite result has be obtained, and the act of measurement (or of observation) which "creates" the result. Indeed, as we have seen, for several reasons, including the contextuality of quantum mechanics, it is not permissible to think that the result of a measurement pre-exists the measurement, or that the event constituted by this measurement occurred spontaneously and that the observer plays a passive role in recording a pre-existing fact. And this is true both for ontic interpretations, where the state vector is supposed to represent reality, and for epistemic representations, where it is merely a representation of information, belief, etc. So let's forget for a moment the question of the state vector and ask ourselves about the occurrence of the event itself.

Some interpretations, such as Healey's pragmatism [16, 17, 18], consider that the fact that the measurement of an observable with several potential results gives a single result is an empirical fact that does not need to be explained. Why and how this result appears is, according to these interpretations, outside the realm of quantum physics. What is more, these interpretations are objective and posit that the result is a fact for the universe (which therefore has the same status as an event in a classical realist framework). In the end, these interpretations place themselves in a realist world that is quite similar to the classical world, the only difference being that objects do not have properties that are all defined at the same time (because of the contextuality) and that when we want to know a given property, the experiment to be carried out to observe it creates the value we are looking for by following a contextual probability calculation. However they do not try to understand how that happens. I don't find this position satisfactory because it casts a veil of secrecy over a mysterious operation, but it's not inconsistent.

As we are going to see below, QBism, on the other hand, postulates that the measurement result is created by the agent's action, that it constitutes his or her personal experience. This is what QBists call participatory realism. How this happens is not made explicit, and although this vagueness is unsatisfactory, the phenomenological framework in which Bitbol tries to place QBism rescues this lack of description. After all, phenomenology tells us that only first-person experience should form the basis of our understanding of the world, and that the way in which this experience is formed is not something we can detail. We can therefore follow QBism revised by Bitbol in this direction and admit that the result is created by the action of the agent who performs a measurement and that this constitutes the agent's personal experience. Where things need to be clarified, however, is when we ask about the status of the result obtained. What is the status of an observer's result for others? Is it a result solely for the agent who made the measurement (as certain statements in QBism seem to indicate) or is it a result which, once obtained by an agent, is imposed on other agents as a fact of the universe? I asked these questions in my review of QBism [25] but none has received a satisfying answer yet.

This is what we analyse in the next section.

## 3. Is QBism perspectival?

QBism can essentially roughly be summed up in two main theses. The first is that it gives primacy to the personal experience of each agent, which is the only thing that is directly accessible. The result of a measurement is then nothing else than the experience of the agent making the measurement. The second is the fact that the quantum formalism (in particular state vectors) does not directly represent reality, but is merely a tool that agents use to calculate the probability of their future experiences, and that is based on a personal (and therefore subjective) estimate of the agent's beliefs as a function of his knowledge of the past. DeBrota and Stacey say [26]:

*A "quantum measurement" is an act that an agent performs on the external world. A "quantum state" is an agent's encoding of her own personal expectations for what she might experience as a consequence of her actions. Moreover, each measurement outcome is a personal event, an experience specific to the agent who incites it.*

Notice that QBists totally endorse the existence of an external world. They say that they stand with Martin Gardner [27]:

*The hypothesis that there is an external world, not dependent on human minds, made of something, is so obviously useful and so strongly confirmed by experience down through the ages that we can say without exaggerating that it is better confirmed than any other empirical hypothesis. So useful is the posit that it is almost impossible for anyone except a madman or a professional metaphysician to comprehend a reason for doubting it.*

DeBrota and Stacey [26] insist on:

*For a QBist, the basic subject matter of quantum theory is an agent's interactions with the outside world; the formalism of quantum theory makes no sense otherwise.*

But quantum mechanics does not directly say something about the "external world". A measurement is just a special case of experience and does not reveal a pre-existing state of affairs but creates a result for the agent. In particular, there is no measurement when there is no agent: a Stern and Gerlach apparatus cannot measure by itself the spin of a particle.

QBism raises several questions that I have presented in a paper [25] to which I refer the reader for a detailed analysis. Here I will analyse mainly the questions which are relative to the ontology.

A first question is to understand the status of this external world. It is difficult to be satisfied with the answer QBists give [26]:

*Accordingly, QBists say that a quantum state is conceptually no more than a probability distribution. Okay, fine, but what is the stuff of the world? QBism is so far mostly silent on this issue, but not because there is no stuff of the world. The character of the stuff is simply not yet understood well enough. Answering this question is the goal, rather than the premise.*

But even if we acknowledge the fact that it is not possible to give precise details about the stuff of the world (it could be the case that it is something of which it is impossible to speak because the language is not suited for that) a first question is: is this world the same for all the agents or is it relative to each agent? According to Mermin [28)]:

*Although I cannot enter your mind to experience your own private perceptions, you can affect my perceptions through language. When I converse with you or read your books and articles in Nature, I plausibly conclude that you are a perceiving being rather like myself, and infer features of your experience. This is how we can arrive at a common understanding of our external worlds, in spite of the privacy of our individual experiences.*

Mermin seems to think that the communication between agents is of a classical nature and speaks as if, beside the actions that agents take to create personal results in quantum experiments as QBism explains, the macroscopic world (language, books, articles in Nature, …) was out of the quantum framework. So it seems that the external world (at the macroscopic level) is the same for all the agents. Nevertheless, in the description of QBism given by Bitbol in the framework of phenomenology, the external world does not seem to "be there" for all the agents but is something personal constructed by each agent for herself:

*At the most fundamental level "the exclusivity of the phenomenon is asserted, that is to say the appearance that emerges from an activity of techno-experimental exploration carried out from within the environment to be explored by self-defined processes of this environment called agents. […] It is up to the agents to construct, if and when they can, the representation of a domain of objects from a well-chosen set of phenomena, and then to represent themselves as those to whom the phenomena appear. The phenomenon is, in one block, the appearance of something*

*to an agent; and the experiment is, in one block, the activity of an agent within something."*. [24, p.463].

QBists are far less clear about their ontology and QBism seen by Bitbol is not exactly the same than QBism seen by QBists… In particular and in accordance with phenomenology Bitbol is more careful than QBists not to postulate too directly the existence of an external world.

So if we want to analyse to which extent QBism is perspectival we have to separate Bitbol's version from the standard QBists' version. The question is to know if an action of an agent providing her a result that constitutes her experience has an impact on the external world that can be part of the experience of another agent or if the experience is purely private. Put it differently, if Alice gets the result "+" when measuring the spin along Oz of a particle, will Bob get the same result if he does the same measurement immediately after? In Bitbol's version it seems that the answer to this question should be that there is no reason for that and even that this question is illegitimate since it can be raised only from a third person point of view (God's point of view) which is not allowed. We are therefore free to act as if the experiences were the same or on the contrary to think that the experiences were different. In this case, the conclusion would be that each agent's experience (including the results she gets and also the external world that she constructs for herself through her perceptions) is totally private and is not something that happens in a common external world.

At the end of his discussion of the arguments of Frauchiger and Renner [29] and of Brukner [30], which show that facts or events are relative to the situation of those who observe them, Bitbol repeatedly insists on reminding us that "lived experience is the only unshakeable certainty, not to say the whole of absolute being (in Husserl's words)". If we follow this statement to the letter, then we must necessarily place ourselves in the position of thinking that experiences of different agents are incommensurable and therefore Bitbol's QBism is fully perspectival. But it is not clear at all in the QBist papers that they are willing to go so far. Mermin's quotation above clearly implies that agents can arrive to a common understanding of their shared external world.

It is true that phenomenology does not preclude the idea that the personal experiences of each individual can give rise to communication between individuals which helps to create at least the illusion of a shared reality. But it remains that this possibility offered by phenomenology is much easier to accept in a classical world where events can be considered objective (even if this is the result of a reconstruction of our perceptions a posteriori) than in a quantum world where, from the outset, it must be assumed that events have a status that is essentially relative to the observer who created them.

As Leifer seems to say in his talk, QBism is not obliged to be perspectival to the maximum level. In this case even though the result gotten by an agent is not objective, it is not totally relative to that agent. The problem is that it is difficult to give a meaning to this. What does that mean to have an impact on a common external world but that this impact is not objective? This halfway position doesn't seem very solid.

We are going to see that Convivial Solipsism draws all the necessary consequences from the observation that the results of observation are entirely relative to the observer, and is therefore a maximally perspectival position.

## 4. Convivial Solipsism (ConSol) ontology in a nutshell

The usual presentation of an interpretation starts from the formalism and gives the meaning of the mathematical components and the way to use them. Then the question of ontology is studied. This is the way I have used in my previous papers on Convivial Solipsism (ConSol here after) [6, 31, 32, 33, 34]. Here I will use the reverse way and will start from the ontology.

Contrary to what its (deliberately provocative) name might suggest, Consol does not in any way defend a solipsistic position. It's even an interpretation that could be described as realist, if we accept to extend the concept of realism in the sense that Kant's transcendental idealism is a realist interpretation, since it posits the concept of "reality in itself".

As I say in a previous paper [31]:

> *Convivial Solipsism is situated in a neo-Kantian framework and assumes that there is "something" else than consciousness, something that (according to the famous Wittgenstein's sentence) it is not appropriate to talk of. This is close to what Kant calls "thing in itself" or "noumenal world".*

I warn the reader that in the beginning of this presentation of the ontology I use colourful terms to try and make the reader understand what I'm talking about, but the language is not adapted to this description and the words used (italicized) should not be taken at face value.

It *happens* that parts of this *something* emerge as systems able to perceive, the conscious observers. So the existence of other observers is also assumed taking ConSol a step further away from solipsism.

To each observer is attached a set of all the potentialities that she could actualize through observations. This set of possibilities is what is called the empirical reality of the observer. The state vector that the observer uses models her empirical reality.

Consciousness and this *something* give rise through observations to what each observer thinks is her reality and that is called in ConSol the phenomenal reality. Following Putnam's famous statement, "the mind and the world jointly make up the mind and the world" [35]. So perceiving is not witnessing what is in front of us but is creating (independently for each us) what we perceive through a co-construction from the world and the mind. In summary, each observers making observations can actualize in her phenomenal reality one of the possibilities that constitute her empirical reality.

A detailed description of what happens in the whole process of perception is of course out of reach. ConSol is close to QBism since it considers also that an observation does not reveal a preexisting state of affairs but is a creation for an individual observer. But where QBism is dumb about how this creation can happen, ConSol explains that each conscious observer builds her own world of which she is aware, her phenomenal reality, by selecting one possible result from her empirical reality modelled by the state vector she is using. This empirical reality should not be seen as an external reality that exists "outside the observer". It is more a field of potentialities that is attached to the observer. Since the empirical reality is relative to each observer, so is the state vector that models it and of course so is the phenomenal reality that each observer creates through personal acts of perception.

It is in fact possible to represent the ontology of ConSol in two ways which, although different in their metaphysical meaning, are rigorously equivalent in terms of what they imply in the resolution of paradoxes.

The first is the one presented above, in which the starting point is this *something* which gives rise to structures capable of perception, the observers. The field of perceptual possibilities of each observer is what we call her empirical reality, which is not considered to exist but is simply the name given to the field of possibilities that she can perceive. This empirical reality is modelled by the state vector that the observer uses to summarize both all her possibilities of observation and the probabilities associated with each result. In this case, empirical reality has the status of a simple name given to the field of possibilities, and the state vector is nothing more than a calculation tool for describing the possibilities of results in the event of observation and the associated probabilities. The fact that the state vector takes the form of a superposition of amplitudes comes from the fact that the probability calculus that is to be used is not a classical calculus but a contextual one. Only phenomenal reality has the status of a kind of reality in that it is what the observer perceives and believes to be her reality.

The second one consists of attributing a more "real" role to the empirical reality and considering that it is the part of this *something* that is the closest to the observer (in a way, the tip of the iceberg) with which the observer can interact through the experiences she makes. This part is described by the superposed state vector and is immutable or, more exactly, it evolves linearly and deterministically according to the Schrödinger equation. This means that it remains entangled. But given the perceptive limits of the observer, it is only possible for her to perceive certain components of this empirical reality, which is not accessible to her in all its superposed richness. Observation therefore consists of taking a look at this empirical reality through an experiment and perceiving a section of this empirical reality which will be considered as the observer's phenomenal reality, whereas the empirical reality will stay completely unaffected.

This second presentation will suit those who wish to rely on an image closer to a certain type of realism. It is the presentation I have often used in previous papers intended to be published in reviews not really open to philosophical abstraction. The empirical reality under this presentation shares many features in common with Bernard d'Espagnat's veiled reality [36]. It is a veiled reality which we can glimpse through our observations, but which is revealed to us only very partially and which our perceptive capacities are powerless to grasp in all its richness.

The first presentation is more radical in the sense that it takes the phenomenological approach to its logical conclusion by practicing the *épochè*, which consists in suspending all judgement concerning the existence of a reality that transcends our perceptions (the empirical reality) beyond the phenomenological reconstruction of a reality that is limited to the minimum field directly resulting from our perceptions (the phenomenal reality).

Whatever presentation is preferred the main feature of ConSol is that it is an interpretation that gives the supremacy to the first person standpoint. The empirical reality and the associated state vector, the results gotten through observations and the phenomenal reality are all strictly private and relative to the observer.

### 5. Problems (dis)solved by Convivial Solipsism

We can now switch to the description of the formalism and the meaning that we should give to it. As we said, the state vector is the way the observer models her empirical reality (i.e. the field of possibilities that she can actualize through observations). It is the tool that both describes what is possible to get as a result of a measurement and which allows to compute the probability to get each result.

Consol rests on two main assumptions. The first one is the hanging-on mechanism which gives a precise definition of what a measurement is (which is missing in the standard interpretation that faces the measurement problem). A measurement is the act of perception we described above that happens when the observer makes an experiment which creates a result for her. The perception of this result is made at random among the different possible results of the corresponding superposed state vector written in the preferred basis[3] and the probability is given by the Born rule. The observer's consciousness hangs-on to the branch corresponding to this result. Once the consciousness is hung-on to one branch, it will hang-on only to branches that are daughters of this branch for all the following observations. That is what is called the hanging-on mechanism. Of course that means that there is no measurement when there is no observer. The interaction between a system and an apparatus cannot give a definite result. It is just an entanglement between the system and the apparatus.

The second one is that state vectors (and all other pieces of the mathematical formalism such as observables or Hamiltonians) are relative to the observer. This is a common point with QBism.

We can see then that the measurement problem does not exist in ConSol since there is a clear definition of what a measurement is and that there is no physical change during a measurement. There is no Schrödinger's cat paradox neither. If we choose the first presentation of the ontology then the superposed state vector of a dead cat and a living cat does not represent anything real but is only a predictive tool allowing to compute the probability to observe a dead cat or a cat alive if the observer looks inside the box. If we choose the second presentation, things are a little bit weirder but we must accept that the empirical reality (if it is supposed to have a kind of existence) does not resemble anything we are accustomed to. In the phenomenal reality which is the only one that is perceived, the cat is either dead or alive.

ConSol assumes that the quantum formalism is universal. It can be used to model not only physical systems of any size but also other observers. That is a very important point. For a given observer the other observers are similar to other physical systems even though they are supposed to be able of perception. The perceptions of an observer are private and no other observer can have any access to them. That means that when Alice makes a measurement and gets a result for her this is not a measurement for Bob. Exactly in the same way as if she was a simple measuring device, when Alice makes a measurement, for Bob, she becomes entangled with the apparatus she uses and the system she wants to measure. This solves the Wigner's friend problem. The fact that the friend made a measurement inside the laboratory and got a result for her does not mean that a measurement has been made for Wigner. So Wigner is totally legitimate to use an entangled state vector to model his friend. From his friend's point of view a measurement has been made and the friend can use a state vector corresponding to a definite result. This is not a problem since we know that state vectors are relative to each observer and we can here perfectly understand why.

---

[3] We use decoherence to define the preferred basis.

Suppose Bob has performed a measurement of the spin along Oz of a spin half particle. According to the hanging-on mechanism, Bob's consciousness is hung-on to one of the two possible branches "+" or "-". But from Alice's point of view, the measurement Bob did is only an interaction between Bob, the apparatus and the system. So Alice attributes an entangled state to the big system [Bob + apparatus + system].

We can now wonder what happens to Alice when she asks Bob which result he got from the measurement he made. Any transfer of information from an observer B to another A – for example, any answer made by Bob to Alice – proceeds through physical means and necessary takes the form of a measurement made by A on B. When she interrogates Bob she makes a measurement on Bob and according to the hanging-on mechanism she can get an answer which is one of the possible Bob's answers (i.e. results that Bob got, "+" or "-").

Now before speaking to Bob Alice can perform the same measurement on the particle and Alice will be hung-on as well to one of the two branches, "+" or "-". Notice that since Bob is entangled with the system and the apparatus, this branch includes the state of Bob that is linked to the very same value. So when Alice, hung-on to that branch, speaks with Bob to know what Bob saw, she performs a measurement on Bob and, in accordance with the hanging-on mechanism, she cannot hear Bob saying anything else than the value that she has got herself. Alice will never hear Bob saying that he saw "+" when she saw "-". No conflict is possible and the intersubjectivity is preserved.

This is why this interpretation is called Convivial Solipsism: each observer lives in her own world but the observers cannot conflict with one another. They necessarily agree. Does that mean that Alice and Bob saw necessarily the same result? This question is meaningless in ConSol.

To avoid falling in false debates, we must emphasize here that ConSol implies that each sentence that is stated has to be attached to one unique observer. Only first person statement are allowed. There is no meta-observer (God's point of view) able to speak simultaneously of the perceptions of two observers. This is the reason why the apparent paradox caused by the question: "how is it possible that Alice hears Bob saying that he saw "+" if, in reality, he saw "-"?" cannot be raised because this question mentions simultaneously what Alice and Bob perceive which is forbidden in ConSol.

We can now notice that in ConSol there is no need of non-locality to explain the EPR paradox. The reason is exactly the same than in QBism. The measurement made by Alice on her particle has absolutely no physical effect on Bob's particle since a measurement has no physical impact. As Bitbol and De la Tremblay say [37]:

> *But if outcomes and predictions are compared in the only place where they can be at the end of the day, namely in the experience of a single agent at a single moment, any contradiction fades away, and even the need for mysterious actions (or passions) at a distance disappears.*

The correlation of results between Alice's measurement and Bob's measurement can only be noticed when Alice meets Bob and asks him which result he got. And this can be done only in a time-like interval.

### 6. Conclusion: Why is Convivial Solipsism maximally perspectival?

As we emphasized above, the empirical reality and the associated state vector, the results gotten through observations and the phenomenal reality are all strictly private and relative to the observer. That means that ConSol is fully compliant with the way Leifer described perspectivalism: "what is true depends on where you are sitting". Now we have to show why it is maximally perspectival contrarily to QBism or at least to the version of QBism that we can assume to be the one defended by its founders.

In ConSol only first person statements are allowed. That implies that each sentence has to be attached to one unique observer; i.e. it must be indexed by an observer. The consequence is that it is meaningless to speak of the possibility that two observers can share a perception, a result or even worse a part of reality. Each observer lives in her own world and there is no commensurability between two observers' worlds. This is why ConSol is maximally perspectival even though it is convivial.

It seems to me that this should also be the logical conclusion that QBism implies. But strangely enough, QBists are not keen to go that far and even though what they say on this subject is far from clear, they seem to be willing to keep the possibility to consider shared knowledge or a shared reality. Mermin's quotation above goes clearly in that direction. Pienaar goes in the same direction [38]:

> *Even more, the extent to which scientific knowledge is 'objective' must be understood as arising from the inter-subjective experiences of many Agents. […] The objective content of quantum theory therefore is not localized in states or measurement results, but rather resides in more holistic structural features of the theory that apply equally to all Agents*

The problem is that this type of description is very fuzzy. It is hard to see how this shared knowledge can emerge from a pure personal experience. Moreover it does not seem to be in agreement with the strong QBist claim made by Fuchs that [39]:

> *This is why QBists opt to say that the outcome of a quantum measurement is a personal experience for the agent gambling upon it. Whereas Bohr always had his classically describable measuring devices mediating between the registration of a measurement's*

> *outcome and the individual agent's experience, for QBism the outcome just is the experience.*

I don't find these different claims very consistent between them and I assume that it is only by shyness that QBists want to conserve a minimum kind of a traditional vision of the world which is not the case for ConSol.

Nevertheless, I must recognize that finding a way to mitigate the radical version of perspectivalism that Consol implies and to account for the fact that we cannot totally reject the idea that there is something that observers share is necessary. That is a work in progress.